\documentclass[aps, prc, showkeys,superscriptaddress, nofootinbib, floatfix]{revtex4-2}
\usepackage{graphicx}
\usepackage{dcolumn}
\usepackage{bm}

\usepackage[bookmarksnumbered, pdfpagelabels=true, plainpages=false, colorlinks=true, linkcolor=blue, citecolor=blue, urlcolor=blue]{hyperref}

\usepackage{ulem}
\usepackage{amssymb}
\usepackage{dsfont}
\usepackage{url}
\usepackage{caption,subcaption}
\usepackage{physics}
\usepackage{xcolor}
\usepackage{geometry}

\newcommand{\be}{\begin{equation}}
\newcommand{\ee}{\end{equation}}
\newcommand{\ba}{\begin{eqnarray}}
\newcommand{\ea}{\end{eqnarray}}
\newcommand{\bd}{\begin{displaymath}}
\newcommand{\ed}{\end{displaymath}}

\def\thalf{{\textstyle{\frac{1}{2}}}}
\def\tthalf{{\textstyle{\frac{3}{2}}}}

\def\twoth{{\textstyle{\frac{2}{3}}}}

\def\ttqt{{\textstyle{\frac{3}{4}}}}

\def\fth{{\textstyle{\frac{4}{3}}}}

\begin{document}

\preprint{APS/123-QED}

\title{Covariant formulation of spinodal decomposition in rapidly expanding quark gluon plasma}

\author{Joseph I. Kapusta}%
\email{kapusta@umn.edu}
\affiliation{School of Physics and Astronomy, University of Minnesota, Minneapolis, MN 55455, USA}

\author{Mayank Singh}
\email{mayank.singh@vanderbilt.edu}
\affiliation{School of Physics and Astronomy, University of Minnesota, Minneapolis, MN 55455, USA}
\affiliation{Department of Physics and Astronomy, Vanderbilt University, Nashville, TN 37240, USA}

\author{Thomas Welle}
\email{twelle2357@gmail.com}
\affiliation{School of Physics and Astronomy, University of Minnesota, Minneapolis, MN 55455, USA}
\affiliation{Applied Research Associates, 8537 Six Forks Road, Raleigh, NC 27615, USA}


\begin{abstract}
Quantum Chromodynamics (QCD) is expected to have a first order phase transition between the confined hadron gas and the deconfined quark gluon plasma at high baryon densities. This will result in phase boundary effects in the metastable and unstable regions. It is important to include these effects in phenomenological models of heavy ion collisions to identify experimental signatures of a phase transition. This requires building intuition on phase separation in rapidly expanding fluids. In this work we present the covariant equations of relativistic hydrodynamics with a phase boundary, provide prescriptions to extend the equation of state to metastable and unstable regions, and show the effects of spinodal separation in a Bjorken flow.

\end{abstract}

\maketitle

\section{Introduction}

Full mapping of the QCD phase diagram has been a long standing challenge since the discovery of QCD five decades ago. The Beam Energy Scan (BES) program at the Relativistic Heavy Ion Collider (RHIC) recorded collisions at different collision energies to explore the QCD phase diagram at a wide range of temperatures and baryon chemical potentials \cite{STAR:2021rls,CPODseries}. Inferring the properties of the phase diagram from the BES data would require careful comparison with the phenomenological models which include all the relevant physics.

Lattice QCD calculations have provided \textit{ab initio} calculations on the zero baryon chemical potential axis of the phase diagram \cite{Aoki:2006we,Ding:2015ona,Arslandok:2023utm}. Unfortunately, lattice QCD cannot be extended to large chemical potentials due to the infamous sign problem. Over the last decade, progress has been made in obtaining the QCD equation of state (EOS) at small baryon chemical potentials $\mu_B$ by expanding the QCD partition function in powers of $\mu_B/T$, where $T$ is the temperature \cite{Hegde:2014sta,Guenther:2017hnx,Bazavov:2017dus,Monnai:2019hkn,Noronha-Hostler:2019ayj,Borsanyi:2021sxv,Kahangirwe:2024cny}. This has been complemented by an EOS estimated from perturbative QCD which can be matched to lattice results at $\mu_B = 0$ \cite{Albright:2014gva}. Recent developments in this area can be found in Ref. \cite{Sorensen:2023zkk}.

The QCD phase diagram is expected to have a first order phase transition at low temperatures and large baryon densities \cite{Fukushima:2010bq,Fukushima:2013rx,Fischer:2018sdj}. Lattice QCD with physical quark masses has shown that there is no true phase transition at zero baryon chemical potential, and that the phase change from a confined hadronic gas to a deconfined quark gluon plasma (QGP) is accomplished by a smooth crossover \cite{Aoki:2006we}. Hence, the first order phase transition line is expected to end somewhere before reaching the zero chemical potential axis. This endpoint is expected to be a second order phase transition critical point in the same universality class as the liquid--gas phase transition and the 3D Ising model. While the exact location of the critical point is not known, once a position is assumed there are different approaches to embed a critical point obeying the right universality properties on a background EOS \cite{Guida:1996ep,Nonaka:2004pg,Parotto:2018pwx,Kapusta:2021oco,Kapusta:2022pny}. Pinpointing the location of critical point will have to rely on testing models with different critical point locations by matching them to experimental data.

Successful isolation of experimental signatures of a critical point and the first order phase transition require implementation of critical fluctuations and the phase boundary effects into the heavy ion simulations. These have proven to be incredibly challenging. There have been some efforts to include the fluctuations in the simulations \cite{Stephanov:1998dy,Stephanov:2008qz,Akamatsu:2016llw,Kapusta:2011gt,Young:2014pka,Singh:2018dpk,An:2019osr,Martinez:2018wia,Sakai:2020pjw,Du:2020bxp,Pihan:2022xcl,Chattopadhyay:2023jfm,Basar:2024qxd}, but a realistic model including critical fluctuations, which can isolate the signatures of critical point has remained elusive. For a summary of these approaches, see Refs. \cite{An:2021wof,Du:2024wjm}.

Simulations have shown that even individual BES collisions span a wide region within the phase diagram and the volume of matter whose trajectory is in the neighbourhood of the critical point is not necessarily very large \cite{Shen:2017ruz,Akamatsu:2018olk,De:2022yxq}. This makes it even more challenging to get clean experimental signature of such a critical region. What may be more realistic to achieve is to detect effects of phase boundary created by a first order phase transition. As this is a wide region on the phase diagram, it is much more plausible that a collision of suitable energy will have a big chunk of its volume passing through the phase transition line. Of course this brings in its own challenges. The phase transition can proceed by two paths: nucleation or spinodal decomposition. Once the system supercools nucleation is the first process to initiate the phase change \cite{nucleate1}.  Generally this will happen in the metastable region.  However, if the nucleation rate is small relative to the expansion rate then the system will enter the unstable region and spinodal decomposition will occur \cite{PhaseFieldReview}. Spinodal decomposition may be initiated by thermal fluctuations or by fluctuations already present in the system.  Most simulations assume the latter and proceed to solve the Cahn-Hilliard equation.  Obviously there are initial state fluctuations present in heavy ion collisions. While the first order phase transition has been extensively studied in chemical and condensed matter systems (for example, see the textbook \cite{provatas}), and has been explored in the non-relativistic limit of low energy nuclear collisions \cite{Skokov:2009yu,Skokov:2010dd}, the relativistic collision systems bring in an additional problem in that it is rapidly expanding to relativistic velocities. There has been some pioneering work on exploring the spinodal region in nuclear collisions \cite{Randrup:2009gp,Randrup:2010ax,Steinheimer:2012gc,Steinheimer:2013gla,Steinheimer:2013xxa,Pratt:2017lce,Steinheimer:2019iso} but, to the best of our knowledge, a covariant energy-momentum tensor including the surface energy has not been written. This is also the first work exploring the effects of phase separation in a fluid rapidly expanding {in the longitudinal direction.

As a first step towards this goal, we write the equations of spinodal decomposition in a rapidly expanding fluid. We provide a prescription to extend the EOS to the metastable and unstable regions, and show the results of numerical simulations of a 1D rapidly expanding QCD medium undergoing the first order phase transition. The thermodynamics of first order phase transition is discussed in Sec. \ref{sec:thermodynamic}. We provide prescriptions to extend the equation of state to metastable regions in Sec. \ref{sec:metastable} and write the equations of hydrodynamics with a phase boundary in Sec. \ref{sec:hydrodynamic}. The equations for the case of Bjorken flow are given in Sec. \ref{section:bjorkenhydro} and we show the effects of first order phase transition on Bjorken fluid in Sec. \ref{sec:numericalBjorken}. We discuss our findings in Sec. \ref{sec:discussion}.

\section{Thermodynamics of a phase transition}\label{sec:thermodynamic}

Consider the Helmholtz free energy functional at temperature $T$
\be
F\{n({\bf x},t)\} = \int d^3x \left[ \thalf K (\bm{\nabla} n)^2 + f(T,n) \right]
= \int d^3x \tilde{f}(T,n).
\ee
The function $f(T,n)$ is the bulk free energy density while $\tilde{f}(T,n)$ includes the gradient term in a Ginzburg-Landau free energy functional.  The gradient term represents the cost in energy in a spatially non-uniform system.  When there are two phases, the value of $K$ is related to the surface energy as described below.  When the density changes by a small amount $\delta n$ the free energy change to first order is
\ba
F\{n + \delta n\} - F\{n \} &=& \int d^3x \left[ K \bm{\nabla} \delta n \cdot \bm{\nabla} n + \frac{\partial f}{\partial n} \delta n \right] \nonumber \\
&=& \int d^3x \left[ K \bm{\nabla} \cdot ( \delta n \bm{\nabla} n) 
- K \delta n \nabla^2 n + \frac{\partial f}{\partial n} \delta n \right].
\ea
The total derivative is converted to a surface integral which vanishes with appropriate boundary conditions.  The local chemical potential is the functional derivative
\be
\tilde{\mu} = \frac{\delta F}{\delta n} = \frac{\partial f}{\partial n} - K \nabla^2 n = \mu - K \nabla^2 n.
\ee
The local isotropic pressure is
\ba\nonumber
\tilde{P} &=& n \mu - [f + \thalf K (\bm{\nabla} n)^2] \\\nonumber
&=& n \frac{\partial f}{\partial n} -f - K n \nabla^2 n - \thalf K (\bm{\nabla} n)^2 \\
&=& P - K n \nabla^2 n - \thalf K (\bm{\nabla} n)^2.
\ea
Assuming negligible dependence of $K$ on temperature the entropy density is as usual
\be
s = - \frac{\partial f}{\partial T},
\ee
and the local energy density is
\be
\tilde{\epsilon} = f - T \frac{\partial f}{\partial T} + \thalf K (\bm{\nabla} n)^2 = \epsilon + \thalf K (\bm{\nabla} n)^2.
\ee
Finally the enthalpy density is
\ba\nonumber
\tilde{w} &=& \tilde{\epsilon} + \tilde{P} = n \frac{\partial f}{\partial n} - T \frac{\partial f}{\partial T} - K n \nabla^2 n \\ &=& \tilde{\mu} n + T s
= w - K n \nabla^2 n.
\ea
This is all consistent with the work of Cahn and Hilliard more than 60 years ago \cite{CahnHilliard1,CahnHilliard2} as well as others, notably by Yang, Fleming and Gibbs \cite{Yang}.

To determine the static density distribution we should minimize the Helmholtz free energy subject to the condition that the number of particles is fixed.  This is done by introducing a Lagrange multiplier $\lambda$ and minimizing the integral
\be
I = \int d^3x \left[ \tilde{f}(T,n) - \lambda n \right].
\ee
The resulting Euler-Lagrange equation is
\be
\partial_i \frac{\partial (\tilde{f} - \lambda n)}{\partial (\partial_i n)} - \frac{\partial (\tilde{f} - \lambda n)}{\partial n} = 0,
\ee
which gives
\be
\lambda = \mu - K \nabla^2 n.
\label{eqEL}
\ee
This means setting $\tilde{\mu} = \lambda$ = constant.  If the system has two phases in equilibrium with each other, and if there is a planar interface between them, then the surface free energy $\sigma$ can be calculated from the well-known formula
\be
\sigma = K \int_{-\infty}^{\infty} dx \left(\frac{dn}{dx}\right)^2,
\ee
where $n(x)$ is the solution to Eq. (\ref{eqEL}).  In that case we can identify $\lambda$ with the common chemical potential of each phase, $\lambda = \mu_L = \mu_G$, along the coexistence curve. 

\section{Metastable and Unstable Regions}\label{sec:metastable}

First principles calculations of the equation of state will usually not describe the metastable and unstable regions of the phase diagram.  Instead, for $T < T_c$ and $n_G \le n \le n_L$ we parameterize, or interpolate, the pressure in terms of the density as
\be\label{eq:eos1}
P_{{\rm int}}(n) = P_X(T) + \sum_{i = 1}^4 c_i (n - n_G)^i.
\ee
The subscript $X$ refers to any point along the line of phase coexistence.  The condition $P_{{\rm int}}(n_L) = P_X(T)$ places one constraint on the four $c_i$
\be\label{eq:eos2}
\sum_{i = 1}^4 c_i \Delta n^i = 0,
\ee
where $\Delta n \equiv n_L - n_G \ge 0$, while continuity of the first derivative at the endpoints provide two more conditions
\ba\label{eq:eos3}
\frac{\partial P}{\partial n}(n_G) &=& c_1 = \frac{n_G}{\chi_G}, \nonumber \\\label{eq:eos4}
\frac{\partial P}{\partial n}(n_L) &=& \sum_{i = 1}^4 i c_i \Delta n^{i-1} = \frac{n_L}{\chi_L}.
\ea
Here $\chi_G$ and $\chi_L$ are the baryon susceptibilities $\chi = n/(\partial P/\partial n)_T$ in the gas and liquid phases.  The chemical potential is 
\be
\mu_{{\rm int}}(n) = \mu_X(T) + d_0 \ln(n/n_G) + \sum_{i = 1}^3 d_i (n-n_G)^i,
\label{chemn}
\ee
where
\ba
d_0 &=& c_1 - 2c_2 n_G + 3 c_3 n_G^2 - 4 c_4 n_G^3, \nonumber \\
d_1 &=& 2c_2 - 3 c_3 n_G + 4 c_4 n_G^2, \nonumber \\
d_2 &=& \tthalf c_3 - 2 c_4 n_G, \nonumber \\\label{eq:eos6}
d_3 &=& \fth c_4.
\ea
This automatically satisfies $\mu_{{\rm int}}(n_G) = \mu_X(T)$.  Equivalently
\ba
c_1 &=& d_0 + d_1 n_G, \nonumber \\
c_2 &=& \thalf d_1 + d_2 n_G, \nonumber \\
c_3 &=& \twoth d_2 + d_3 n_G, \nonumber \\
c_4 &=& \ttqt d_3.
\ea
The final condition $\mu_{{\rm int}}(n_L) = \mu_X(T)$ is satisfied when
\be\label{eq:eos7}
d_0 \ln(n_L/n_G) + \sum_{i = 1}^3 d_i \Delta n^i = 0.
\ee
Then all coefficients $c_i$ and $d_i$ are determined. 

To test the veracity of this parameterization we compare to the Van der Waals equation of state.  Its dimensionless form, representing the law of corresponding states, is
\be
P_* = \frac{8T_*}{3v_* - 1} - \frac{3}{v_*^2}.
\ee
Here $P_* = P/P_c$, $v_* = v/v_c$, $T_* = T/T_c$, and $v = 1/n$.  Some results are shown in Fig. \ref{fig:VdW}.  
\begin{figure}
    \centering
    \includegraphics[width=0.45\textwidth]{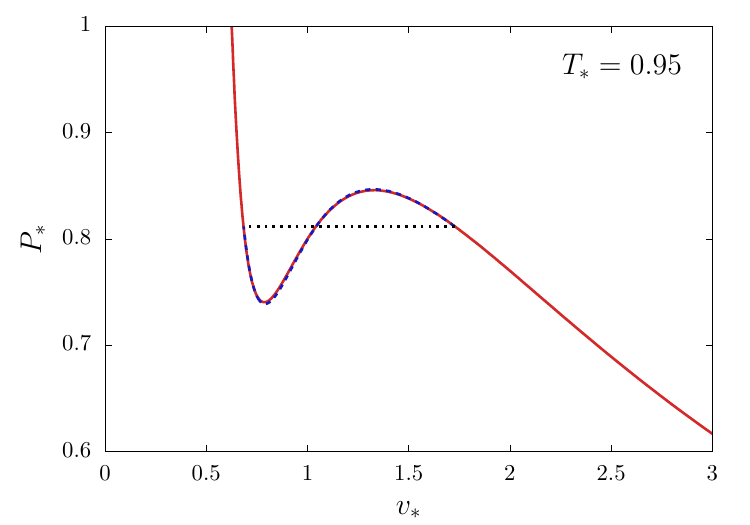}
    \includegraphics[width=0.45\textwidth]{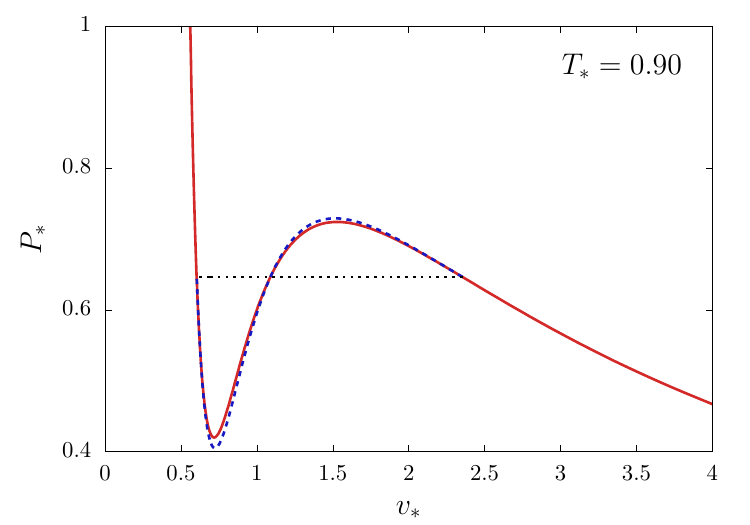}
    \includegraphics[width=0.45\textwidth]{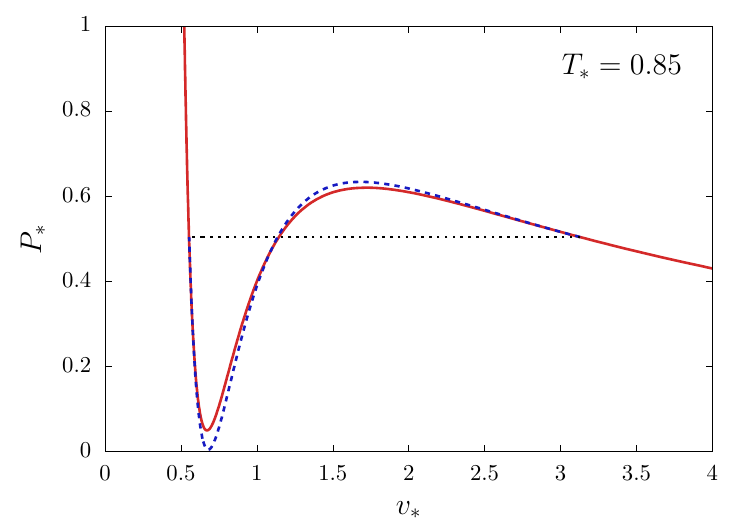}
    \includegraphics[width=0.45\textwidth]{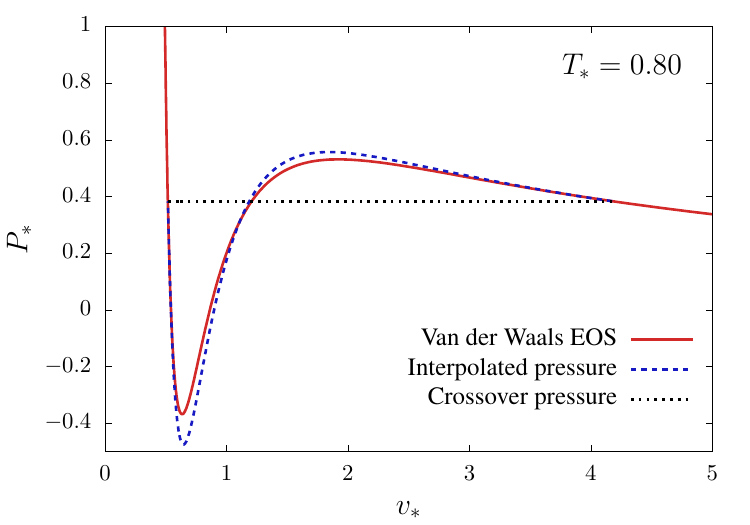}
    \caption{The parameterization of the metastable and unstable regions, as described in the text, compared to the Van der Waals equation of state.  The comparison is done for $T_*$ = 0.95, 0.90, 0.85, and 0.80.  The plot shows the volume per particle even though the parameterization was done in terms of the density $n_* = 1/v_*$ to illustrate the Maxwell construction.}
    \label{fig:VdW}
\end{figure}
Alternatively, the pressure can be expanded in terms of the volume per baryon $v = 1/n$. We do not use this construction as we found that the agreement with the Van der Waals equation of state is not nearly as good.  Note that the pressure goes negative in the metastable and unstable regions at low temperature.  This can be handled by the theory presented herein.

For illustration, in this paper we use the QCD equation of state from Ref. \cite{Kapusta:2021oco}. The critical point is embedded on a smooth background equation of state such that the right critical exponents from the 3D Ising model universality class are obtained. The background equation of state is obtained by matching a perturbative-QCD EOS in the quark gluon phase with a hadron resonance gas EOS in the hadronic phase with a smooth matching function \cite{Albright:2014gva}.  The critical point is chosen to be at $T = 130$ MeV and $\mu_B = 450$ MeV. The resulting interpolated chemical potential and pressure at a fixed temperature are shown in Fig. \ref{fig:EOS}.
\begin{figure}
    \centering
    \includegraphics[width=0.45\textwidth]{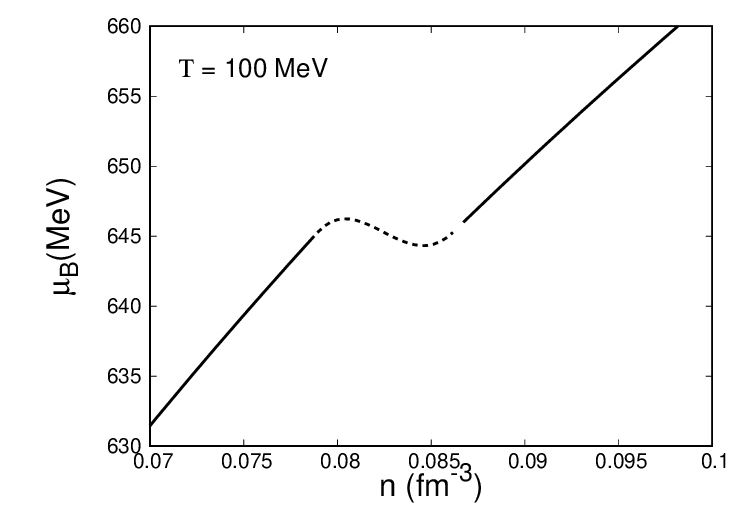}
    \includegraphics[width=0.45\textwidth]{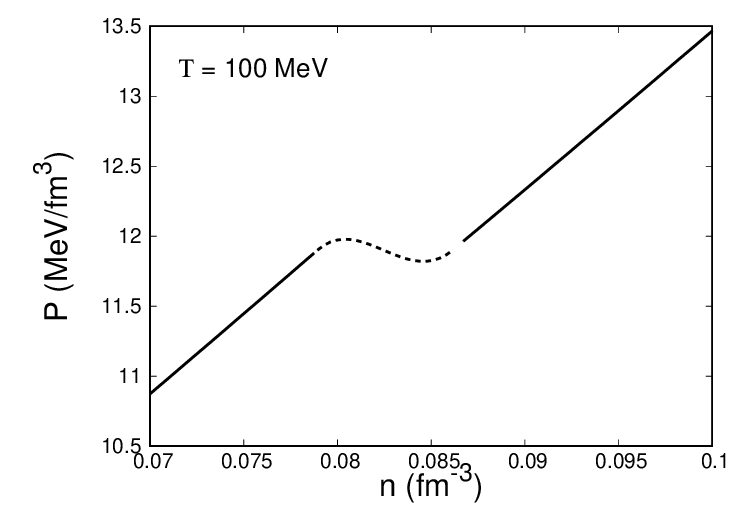}
    \caption{The chemical potential (left) and pressure (right) as a function of baryon density at temperature of 100 MeV. }
    \label{fig:EOS}
\end{figure}

It must be pointed out that the interpolation given above is relevant for an equation of state with mean field critical exponents.  The QCD equation of state employed here uses the known critical exponents from the 3D Ising model universality class.  It should not make much difference except when one is in the neighborhood of the critical point.  In this work we stay away from that neighborhood, focusing instead on crossing the line of first order phase transition during a heavy ion collision.  Generalization of the polynomial interpolation is presented in the appendix.

\section{Hydrodynamics: General}\label{sec:hydrodynamic}

The stress tensor was derived in Ref. \cite{Yang} to be
\be
T_{ij} = \tilde{P} \delta_{ij} + K (\partial_i n) (\partial_j n),
\ee
which is symmetric.  This is in a stationary fluid or gas.  A fully relativistic expression for the stress--energy--momentum tensor is
\be
T^{\mu\nu} = \tilde{P} (u^{\mu} u^{\nu} - g^{\mu\nu}) + \tilde{\epsilon} u^{\mu} u^{\nu} + K (D^{\mu} n) (D^{\nu} n),
\ee
where $u^{\mu}$ is the local flow velocity, the metric is $(+---)$, and
\be
D^{\mu} n \equiv \partial^{\mu} n - u^{\mu} u^{\alpha} \partial_{\alpha} n 
\ee
is a gradient orthogonal to the flow velocity.  This $T^{\mu\nu}$ is obviously a symmetric second rank tensor.  In the local rest frame 
\ba
T^{00} &=& \tilde{\epsilon}, \nonumber \\
T^{0j} &=& 0, \nonumber \\
T^{ij} &=& \tilde{P} \delta_{ij} + K (\partial_i n) (\partial_j n).
\ea
As a check, consider the equation of motion $\partial_{\mu} T^{\mu\nu} = 0$ for a planar interface between the liquid and gas phases with $n({\bf x}, t) = n(x)$.  The only nontrivial equation is
\be
\frac{d}{dx} \left[ \tilde{P} + K \left(\frac{dn}{dx}\right)^2 \right] = 0.
\ee
Substituting for $\tilde{P}$ this can be written as
\be
n \tilde{\mu} - f + \thalf K \left(\frac{dn}{dx}\right)^2 = {\rm constant}.
\ee
This is easily shown \cite{Yang} to be equivalent to Eq. (\ref{eqEL}).

The local thermodynanic variables can be written covariantly using $D^{\mu}$.
\ba
\tilde{\mu} &=& \mu + K D^2 n, \nonumber \\
\tilde{P} &=& P + K n D^2 n + \thalf K (D^{\mu} n) (D_{\mu} n), \nonumber \\
\tilde{\epsilon} &=& \epsilon - \thalf K (D^{\mu} n) (D_{\mu} n), \nonumber \\
\tilde{w} &=& \tilde{P} + \tilde{\epsilon} = Ts + \tilde{\mu} n = w + K n D^2 n.
\ea

\section{Hydrodynamics: Bjorken}\label{section:bjorkenhydro}

In boost-invariant hydrodynamics \cite{Cooper:1974qi,Chiu:1975hw,Bjorken:1982qr} one expresses the time and location along the beam direction in terms of the proper time $\tau$ and space-time rapidity $\xi$ as
\ba
t &=& \tau \cosh\xi, \nonumber \\
z &=& \tau \sinh\xi,
\ea
with the inverse relations
\ba
\tau &=& \sqrt{t^2 - z^2}, \nonumber \\
\xi &=& \tanh^{-1}(z/t)  .
\ea
The flow velocity has the non-vanishing components
\ba
u^0 &=& \cosh\xi, \nonumber \\
u^3 &=& \sinh\xi. 
\ea
It is useful to note that
\ba
\frac{\partial}{\partial t} &=& \cosh\xi \frac{\partial}{\partial \tau} - \frac{\sinh\xi}{\tau} \frac{\partial}{\partial \xi}, \nonumber \\
\frac{\partial}{\partial z} &=&  -\sinh\xi \frac{\partial}{\partial \tau} + \frac{\cosh\xi}{\tau} \frac{\partial}{\partial \xi}, \nonumber \\
\partial \cdot u &=& \frac{1}{\tau}, \nonumber \\
u \cdot \partial &=& \frac{\partial}{\partial \tau}, \nonumber \\
D^0 &=& -\frac{\sinh\xi}{\tau} \frac{\partial}{\partial \xi}, \nonumber \\
D^3 &=& -\frac{\cosh\xi}{\tau} \frac{\partial}{\partial \xi}. 
\ea

Baryon number conservation $\partial_{\mu} (n u^{\mu}) = 0$ quickly leads to
\be
\frac{dn}{d\tau} + \frac{n}{\tau} = 0.
\label{BJn}
\ee
Due to boost invariance and the assumption of transverse homogeneity in the Bjorken model, $D^{\mu}n = 0$ so that energy and momentum conservation $\partial_{\mu} T^{\mu\nu} = 0$ lead to
\be
\frac{d\epsilon}{d\tau} + \frac{w}{\tau} = 0.
\ee
At fixed volume $d\epsilon = T ds + \mu dn$.  Together with Eq. (\ref{BJn}) this yields the familiar equation for entropy conservation
\be
\frac{ds}{d\tau} + \frac{s}{\tau} = 0.
\label{BJs}
\ee
All effects of the spatial gradient terms with coefficient $K$ drop out due to the assumption of boost invariance and transverse homogeneity.

Numerical studies of spinodal decomposition typically use a diffusion equation.  In the Landau-Lifshitz definition of flow velocity the baryon current is
\be
J^{\mu} = n u^{\mu} + \sigma_B T D^{\mu} \left( \frac{\tilde{\mu}}{T} \right),
\ee  
where $\sigma_B$ is the baryon conductivity.  This vanishes under the assumption of boost invariance and transverse homogeneity.  The Cattaneo expression for the current is \cite{Cattaneo1,Cattaneo2,KapustaYoung}
\be
J^{\mu} = n u^{\mu} + \sigma_B T D^{\mu} (1 + \tau_B u \cdot \partial)^{-1} \left( \frac{\tilde{\mu}}{T} \right),
\ee
where $\tau_B$ is the relaxation time.  This expression vanishes under boost invariance and transverse homogeneity as well. The Cattaneo expression is causal under linear fluctuations while the diffusion equation suffers from acausality even under small perturbations. Nevertheless, we use the diffusion equation in this work for better correspondence with the non-relativistic literature. The expressions can be changed to their causal version using the Cattaneo form in a straightforward manner.

Diffusion is necessary for spinodal decomposition, or phase separation, as it facilitates the flux from lower to higher density region when the chemical potential gradient is opposite to the density gradient in the unstable region. So we should allow the thermodynamic functions to depend on $\xi$ as well as $\tau$. The energy-momentum conservation equation is

\ba
\partial_{\mu} T^{\mu\nu} &=& \left( \frac{\partial \tilde{w}}{\partial \tau} + \frac{\tilde{w}}{\tau} \right) u^{\nu} - g^{\mu\nu} \partial_{\mu} \tilde{P}  + K \left[ (D^{\nu}n) (\partial_{\mu} D^{\mu} n) + (D^{\mu}n) (\partial_{\mu} D^{\nu} n) \right] = 0.
\ea
In what follows we allow the baryon density to depend on $\xi$ but the energy density and pressure are taken to be boost invariant. As the system evolves, the baryon density variation will break the boost invariance of the energy density and pressure which would lead to an acceleration term. In this work we treat this as a perturbation and neglect the feedback. As we shall see, this is justified in our numerical example.  The implications of a fully boost-invariance breaking general system of equations is left for future work. Some useful expressions are
\be
\partial_{\mu} D^{\mu} = - \frac{1}{\tau^2} \frac{\partial^2}{\partial \xi^2} = D_{\mu} D^{\mu}.
\ee
Energy conservation is
\ba
\left( \frac{\partial \tilde{\epsilon}}{\partial \tau} + \frac{\tilde{w}}{\tau} \right) \cosh\xi + \frac{\sinh\xi}{\tau} \frac{\partial \tilde{P}}{\partial \xi} 
+ \frac{K}{\tau^3} \left[ 2 \sinh\xi \frac{\partial^2 n}{\partial \xi^2} + \cosh\xi \frac{\partial n}{\partial \xi} \right]  \frac{\partial n}{\partial \xi} &=& 0.
\ea
and momentum conservation is
\ba\nonumber
\left( \frac{\partial \tilde{\epsilon}}{\partial \tau} + \frac{\tilde{w}}{\tau} \right) \sinh\xi + \frac{\cosh\xi}{\tau} \frac{\partial \tilde{P}}{\partial \xi} 
+ \frac{K}{\tau^3} \left[ 2 \cosh\xi \frac{\partial^2 n}{\partial \xi^2} + \sinh\xi \frac{\partial n}{\partial \xi} \right]  \frac{\partial n}{\partial \xi} &=& 0.
\ea
Linear combinations yield 
\be
\frac{\partial \tilde{\epsilon}}{\partial \tau} + \frac{\tilde{w}}{\tau} + \frac{K}{\tau^3} \left( \frac{\partial n}{\partial \xi} \right)^2 = 0,
\label{tildee}
\ee
and
\be
\frac{\partial \tilde{P}}{\partial \xi}  + \frac{2K}{\tau^2} \frac{\partial n}{\partial \xi} \frac{\partial^2 n}{\partial \xi^2}  = 0,
\label{tildep}
\ee
where
\be
\tilde{P} = P - \frac{K}{\tau^2} n \frac{\partial^2 n}{\partial \xi^2}
- \frac{K}{2\tau^2} \left( \frac{\partial n}{\partial \xi} \right)^2.
\ee
Expressed in terms of the local thermodynamic functions they are
\be
\frac{\partial \epsilon}{\partial \tau} + \frac{w}{\tau} + \frac{K}{\tau^2} \frac{\partial n}{\partial \xi}  \frac{\partial^2 n}{\partial \tau \partial \xi} 
- \frac{K}{\tau^3} n \frac{\partial^2 n}{\partial \xi^2} = 0,
\label{e}
\ee
and
\be
\frac{\partial P}{\partial \xi} - \frac{K}{\tau^2} n \frac{\partial^3 n}{\partial \xi^3}  = 0.
\label{p}
\ee
Differentiating Eq. (\ref{e}) with respect to $\xi$ and subtracting Eq. (\ref{p}) leads to
\be
\frac{\partial}{\partial \tau} \frac{\partial}{\partial \xi} \left[ \tau \epsilon + \frac{K}{2\tau} \left( \frac{\partial n}{\partial \xi} \right)^2 \right] = 0.
\ee
Now consider baryon number conservation $\partial_{\mu} J^{\mu} = 0$.  For ordinary diffusion one finds
\be
\frac{\partial}{\partial \tau} (\tau n) - \frac{\sigma_B T}{\tau} \frac{\partial^2}{\partial \xi^2} \left( \frac{\tilde{\mu}}{T} \right)
- \frac{1}{\tau} \frac{\partial}{\partial \xi} (\sigma_B T) \frac{\partial}{\partial \xi} \left( \frac{\tilde{\mu}}{T} \right) = 0.
\ee
Note that
\be
\tilde{\mu} = \mu - \frac{K}{\tau^2} \frac{\partial^2 n}{\partial \xi^2}.
\ee

The $K$ is determined by the surface free energy.  We take it to be a constant.  We use the simplified expression for baryon conductivity \cite{Shen:2017ruz}
\be
\sigma_B = \frac{1}{3}\frac{C_B}{T}n\left[\coth\left(\frac{\mu_B}{T}\right) -\frac{3Tn}{w}\right],
\ee
where $C_B$ is a dimensionless number. We have taken its value to be 0.5. A more rigorous expression for $\sigma_B$ and its comparison with the above form is discussed in \cite{De:2022yxq,JuanJoe}.

In order to avoid solving partial differential equations with derivatives greater than two we could choose $\epsilon(\tau, \xi)$ and $n(\tau,\xi)$ as the independent thermodynamic variables.  Then $P$, $T$, $w$, $s$, and $\mu$ are determined via the equation of state $f(n,T)$.

The equations that we solve are
\be\label{eq:solve1}
\frac{\partial \epsilon(n,T)}{\partial \tau} + \frac{w(n,T)}{\tau} + \frac{K}{\tau^2} \frac{\partial n}{\partial \xi}  \frac{\partial^2 n}{\partial \tau \partial \xi} 
- \frac{K}{\tau^3} n \frac{\partial^2 n}{\partial \xi^2} = 0,
\ee
\be\label{eq:solve2}
\frac{\partial}{\partial \tau} (\tau n) - \frac{\sigma_B T}{\tau} \frac{\partial^2}{\partial \xi^2} \left( \frac{\tilde{\mu}}{T} \right)
- \frac{1}{\tau} \frac{\partial}{\partial \xi} (\sigma_B T) \frac{\partial}{\partial \xi} \left( \frac{\tilde{\mu}}{T} \right) = 0,
\ee
\be\label{eq:solve3}
\tilde{\mu} = \mu(n,T) - \frac{K}{\tau^2} \frac{\partial^2 n}{\partial \xi^2}.
\ee
\begin{figure}
    \centering
    \includegraphics[width=0.6\textwidth]{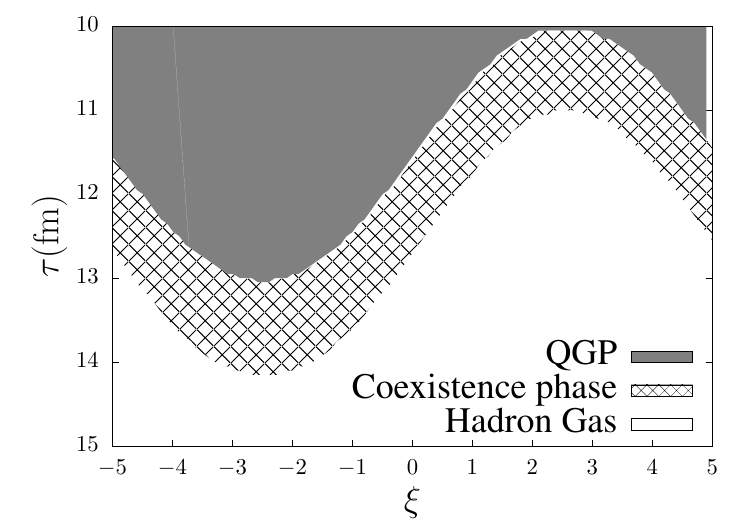}
\caption{Phase of the system as a function of rapidity and time.}
    \label{fig:phase}
\end{figure}

\begin{figure}
    \centering
    \includegraphics[width=0.6\textwidth]{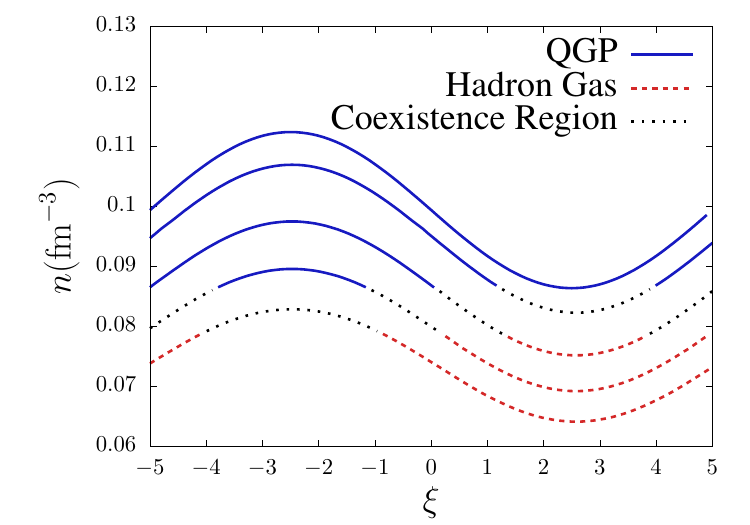}
    \caption{Baryon density evolution at different times. Proper times for curves, from top to bottom, are 10.0 fm, 10.5 fm, 11.5 fm, 12.5 fm and 13.5 fm}
    \label{fig:baryon_density}
\end{figure}

\section{Spinodal separation in Bjorken hydrodynamics}\label{sec:numericalBjorken}

We initialize the system at time $\tau_x = 10$ fm such that the energy density $\epsilon(\tau_x)$ is $\xi$-independent while the baryon density has the periodic fluctuation $n_{fl} \sin(k \xi)$ around the average value $n(\tau_x)$. This is obviously a simplification. However, it is motivated by the fact that while the energy density is boost invariant around mid-rapidity in most initial state models, the baryon density is not \cite{De:2022yxq,Shen:2017bsr}. The fluctuation is necessary to seed spinodal decomposition. These are chosen such that the system just reaches the coexistence curve. Specifically, the point $\epsilon(\tau_x)$, $n(\tau_x) - n_{fl}$ lies on the coexistence curve. We evolve the system following Eqs. (\ref{eq:solve1}-\ref{eq:solve3}).

A periodic boundary condition is used in the $\xi$ direction to preserve boost invariance. The surface energy is proportional to the coefficient $K$. The surface energy term builds up the energy in the coexistence region and causes spontaneous separation of phases as the system approaches equilibrium. This has been a topic of extensive study in the phase dynamics community. Our system has a unique feature in that it is expanding rapidly. As the system expands and cools, the phase change is guaranteed and the system which starts with all QGP will eventually end up as all hadron gas. But different parts of the system will undergo the phase transition at different times. Figure \ref{fig:phase} shows the evolution of the phases for our system at different spacetime points.

The first order phase transition will imprint its signature on the correlations of hadrons. More specifically, it will amplify the effects of local fluctuations which may be seeded at the initial stage or may arise dynamically. To study the effects of the surface energy terms on the system, we solve the system with $K = 0$ and 
$K = 5\times10^{-5}$ MeV$^{-4} \approx 1.5\times10^{7}$ MeV fm$^5$.

\begin{figure}
    \centering
    \includegraphics[width=0.45\textwidth]{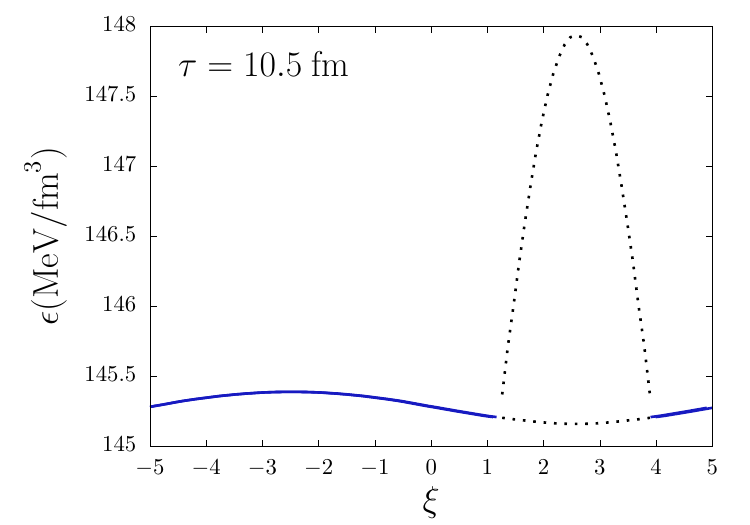}
    \includegraphics[width=0.45\textwidth]{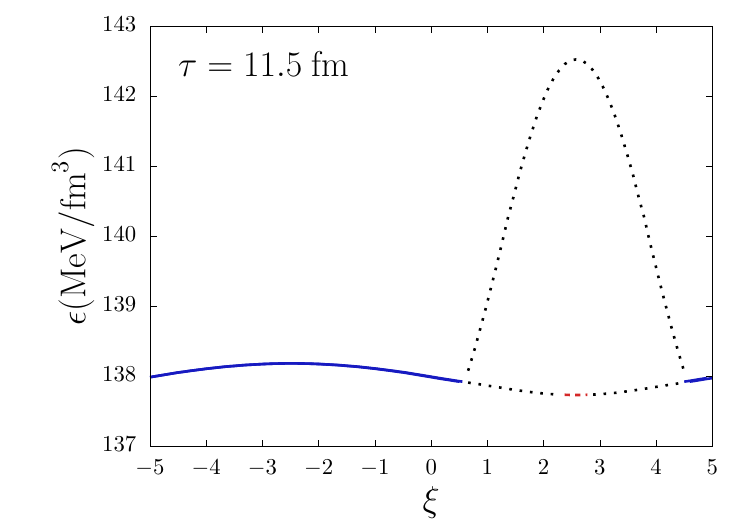}
    \includegraphics[width=0.45\textwidth]{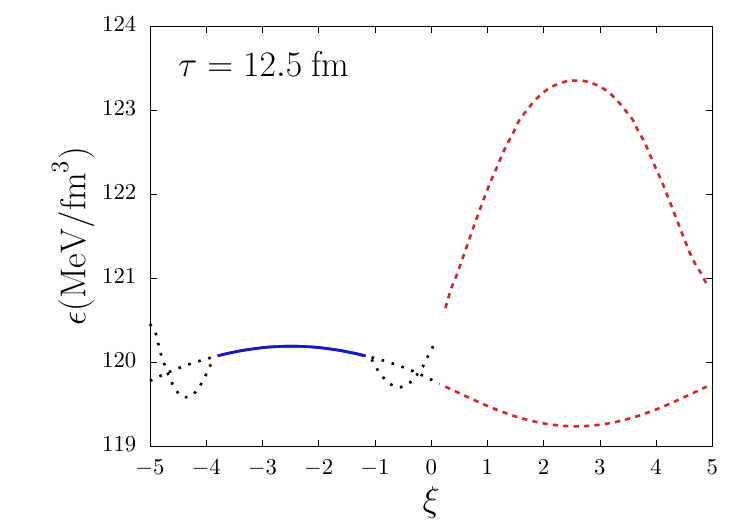}
    \includegraphics[width=0.45\textwidth]{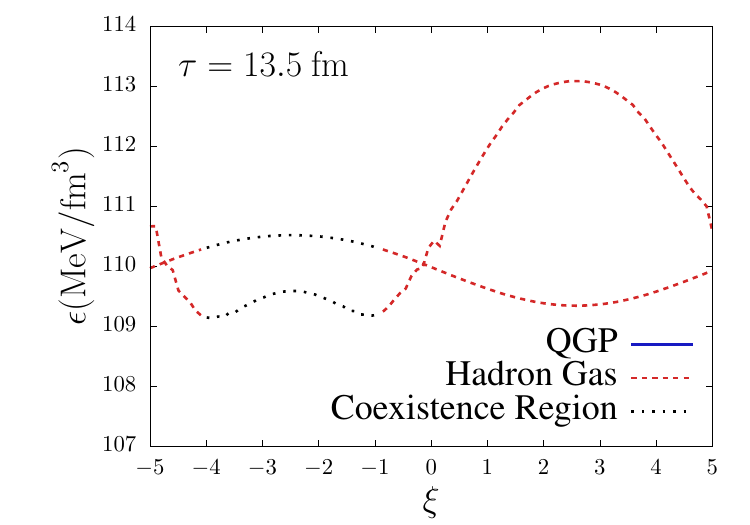}
    \caption{Energy density profile at different times. The curve with the highest energy density point corresponds to $K = 5 \times 10^{-5}$ MeV$^{-4}$ while the other curve has $K = 0$. The energy density in the QGP phase can be lower than the energy density in the coexistence region or in the hadron gas phase because of the gradients of baryon density.}
    \label{fig:energy_density}
\end{figure}

The baryon density evolution is primarily governed by the expansion in the longitudinal direction and by diffusion. The surface term modifies the baryon density evolution through the modification of the the chemical potential in Eq. (\ref{eq:solve3}). This modification is proportional to the fourth derivative in baryon density and is very sensitive to local fluctuations. A small nucleus of matter in the coexistence region can trigger the spinodal decomposition. Due to the fourth order derivative playing a vital role here, it is incredibly difficult to model this using finite-difference methods and a realistic heavy-ion simulation attempting to incorporate first-order phase transition physics will probably need to resort to finite-element methods tailored for this system. Here, we have side-stepped this problem by choosing a smoothly varying baryon density profile. Figure \ref{fig:baryon_density} shows the baryon density evolution for $K=0$. We checked that the baryon density evolution is essentially identical for different values of $K$ when $K \lesssim 10^{-4}$ MeV$^{-4}$. This ensures that the gradients in $n$ are governed by the initial sinusoidal profile and not the small fluctuations making the numerics simpler. As discussed above, this will most likely not be the case in a realistic setting where more advanced techniques will be necessary.

The energy contribution from the surface terms is dependent on the first, second and third order spatial derivatives of the baryon density in the coexistence phase (see Eq. (\ref{eq:solve1})). As baryon density is smoothly varying and the surface correction to it are small, we can numerically evaluate it with good accuracy. The surface energy contribution is not negligible here for the chosen values of $K$. Figure \ref{fig:energy_density} shown the evolution of energy density. We see an immediate energy build up as the system enters the coexistence phase. The energy evolution in the QGP phase is unaffected for different choices of $K$, as it has not yet seen the phase boundary. The energy ordering is inverted in the negative rapidity region at $\tau=13.5$ fm. This is because of the opposite sign of the baryon density spatial derivatives when this region reaches the coexistence phase.

We can see the variations in temperature and chemical potential as a consequence of the phase transition in Figs. \ref{fig:temperature} and \ref{fig:mub}. It is interesting to note that in this particular case, the hadron gas has higher temperature than the QGP phase. This is because of our choice of the initial state where we took a constant energy density and a spatially varying baryon density, hence the phase is being governed by baryon density as the change in energy density is relatively small. For a constant energy density, the region with lower baryon density has higher temperature. Since the energy density variation is small, temperatures are higher where baryon densities are lower and so QGP is cooler than the hadron gas. On the other hand, the baryon chemical potential is higher for QGP than for the hadron gas.

\begin{figure}
    \centering
    \includegraphics[width=0.45\textwidth]{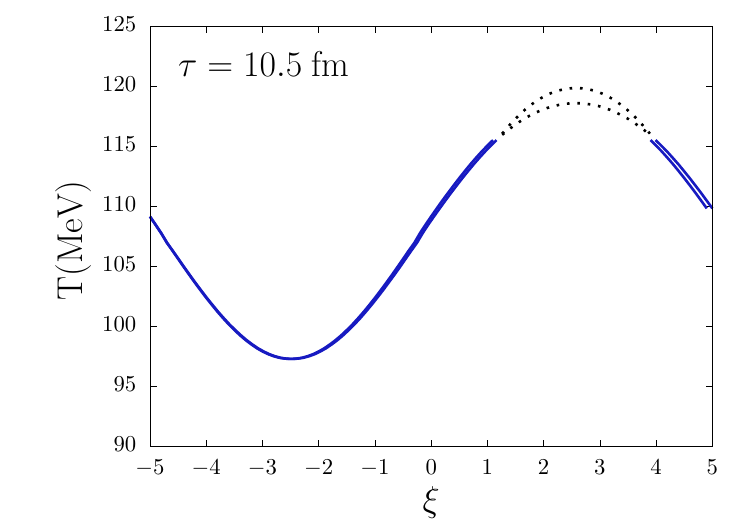}
    \includegraphics[width=0.45\textwidth]{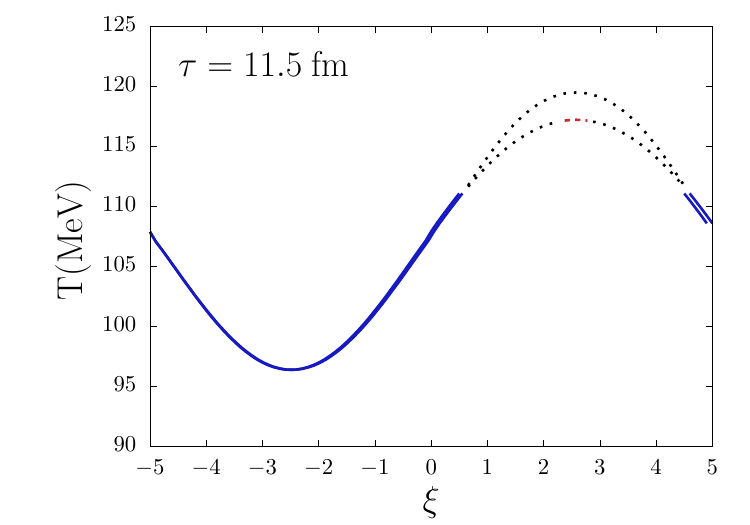}
    \includegraphics[width=0.45\textwidth]{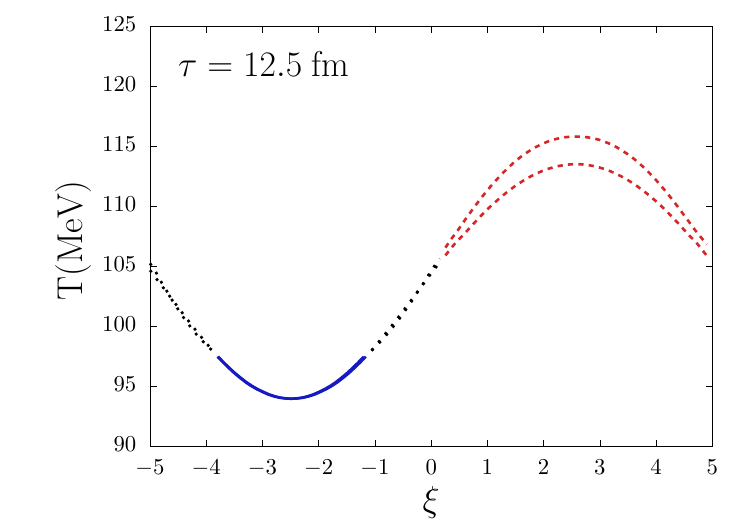}
    \includegraphics[width=0.45\textwidth]{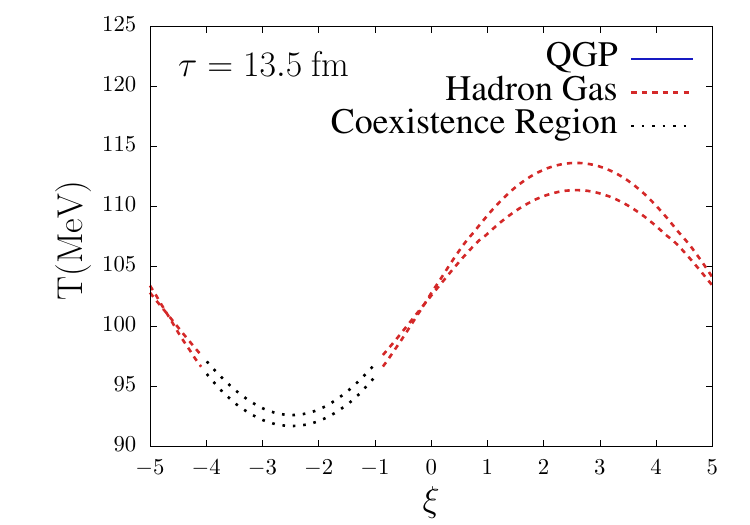}
    \caption{Temperature profile at different times. The curve with the highest temperature point corresponds to $K = 5 \times 10^{-5}$ MeV$^{-4}$ while the other curve has $K = 0$.}
    \label{fig:temperature}
\end{figure}

\begin{figure}
    \centering
    \includegraphics[width=0.45\textwidth]{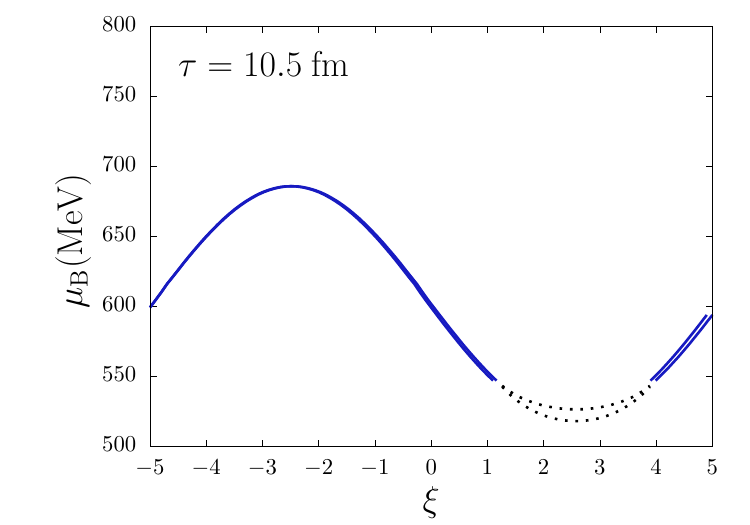}
    \includegraphics[width=0.45\textwidth]{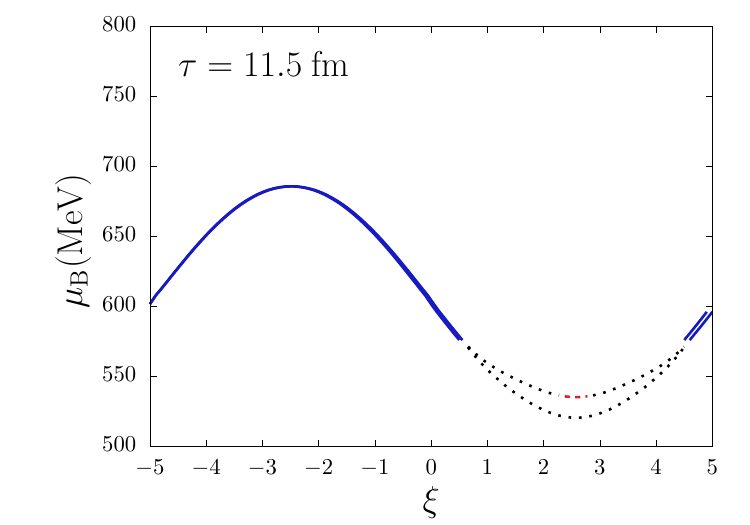}
    \includegraphics[width=0.45\textwidth]{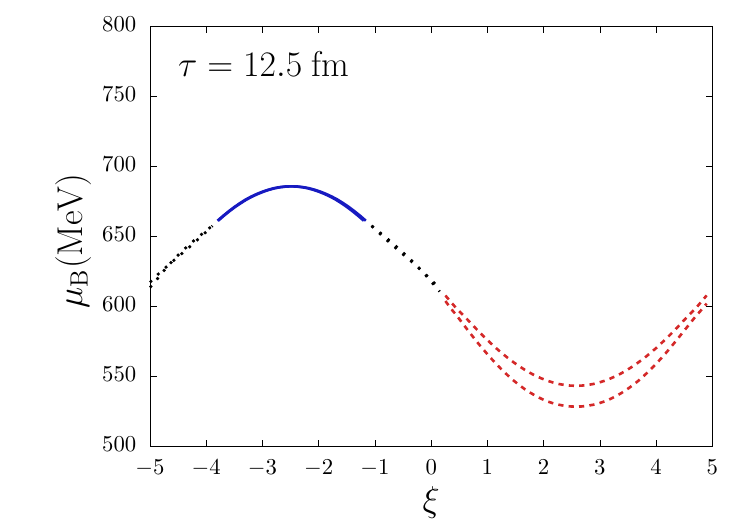}
    \includegraphics[width=0.45\textwidth]{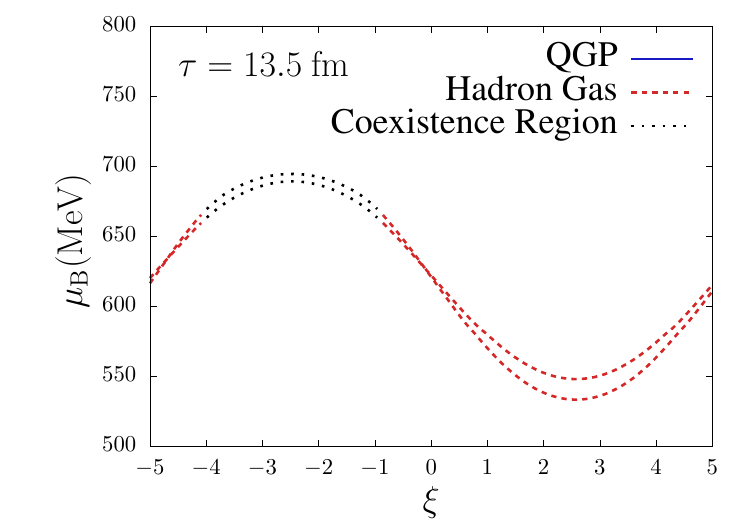}
    \caption{Baryon chemical potential profile at different times. The curve with the lowest baryon chemical potential point corresponds to $K = 5 \times 10^{-5}$ MeV$^{-4}$ while the other curve has $K = 0$.}
    \label{fig:mub}
\end{figure}

\section{Discussion}\label{sec:discussion}

Numerous and various model calculations have fueled speculation that QCD exhibits a line of first order phase transition in the temperature versus baryon chemical potential phase diagram, ending in a critical point at some $T_c$ and $\mu_c$, which is in the same universality class as the 3D Ising model and the liquid-gas phase transition.  Due to the fermion sign problem, it is not known (yet) whether lattice QCD confirms or denies this purported phase structure.  In any event, physics is an experimental science, and the only conceivable way to test this hypothesis in terrestrial experiments is with relativistic heavy ion collisions.  The Standard Model of these collisions includes a stage using second order, relativistic hydrodynamics, which naturally needs an equation of state.  If the transition is slow and adiabatic, the Maxwell construction may be used.  If the transition is faster, then nucleation of hadronic bubbles in the quark-gluon plasma may provide a better quantitative description.  This description applies when the system super-cools into the metastable region.  When the system expands and cools very rapidly, as is likely the case in heavy ion collisions, a more detailed description may be necessary.  This is commonly referred to as spinodal decomposition, and is relevant even in the unstable region.  Spinodal decomposition dynamics has been extensively studied in several branches of physics and chemistry, but only for systems with fixed volume and temperature.  The dynamics in heavy ion collisions is much more challenging.

In this manuscript we have incorporated terms into the stress-energy-momentum tensor, which are normally included to describe spinodal decomposition in non-relativistic systems, into covariant hydrodynamics.  These terms are second-order in thermodynamic densities, namely, the baryon density and its space-time derivatives.  Only one new parameter, $K$, enters and is associated with the surface free energy between the two phases.  It is also necessary to have a parameterization of the equation of state when the system is in the metastable or unstable regions.  This is better accomplished by expanding in powers of the baryon density rather than the volume per baryon.  The resulting equations of motion are fourth order in space-time derivatives.  This makes solution of the space-time evolution very numerically challenging. Therefore, in this manuscript we only gave a simple illustration of how this works in a nearly boost-invariant Bjorken scenario. Motivated by various initial state models, we have used the simplification that the energy density is boost-invariant while the baryon density fluctuates. Once the system is initialized this way, the violation of boost invariance of the energy density is less than 3\% for the system considered here. The pressure gradient terms are ignored. If they were included, the violation of boost invariance would likely be even smaller as those terms would further equilibrate the system. A small value of K was chosen to avoid breaking the smoothly varying nature of the baryon density. Future work should include realistic initial state, hydrodynamic fluctuations, and any other relevant terms.  Only then will comparisons of theoretical simulations with experimental observations fulfill our goal.  Given advances in computational calculations over the past two decades we are optimistic that the goals can be achieved.



\begin{acknowledgments}
We thank U. Heinz and C. Gale for constructive comments on the manuscript.  This work was supported by the U.S. DOE Grants No. DE-FG02-87ER40328 (JIK, MS, and TW) and  DE-SC-0024347 (MS).
\end{acknowledgments}

\section*{appendix}

Parameterization of the metastable and unstable regions in Sec. \ref{sec:metastable} is consistent with mean field values of the critical exponents, specifically $\delta = 3$. For systems in the same universality class as the 3D Ising model and the liquid-gas phase transition the more accurate value is $\delta \approx 4.79$.  In that case we write \cite{Kapusta2}
\be
P_{{\rm int}}(n) = P_X(T) + a_1 (n - n_G) + a_2 (n - n_G)^2 + a_3 (n - n_G)^{\delta - 1} + a_4 (n - n_G)^{\delta}
+ a_5 (n - n_G)^{\delta + 1}
\ee
and
\be
\mu_{{\rm int}}(n) = \mu_X(T) + b_0 \ln(n/n_G) + b_1 (n - n_G) + b_2 (n - n_G)^{\delta - 1} + b_3 (n - n_G)^{\delta}
\ee
Note the degeneracy when $\delta = 3$.
At fixed $T$
\be
n \frac{\partial \mu_{\rm int}}{\partial n} = \frac{\partial P_{\rm int}}{\partial n}
\ee
This leads to
\ba
a_1 &=& b_0 + b_1 n_G \nonumber \\
a_2 &=& \thalf b_1 \nonumber \\
a_3 &=& b_2 n_G \nonumber \\
a_4 &=& \left( \frac{\delta - 1}{\delta} \right) b_2 + b_3 n_G\nonumber \\
a_5 &=& \left( \frac{\delta}{\delta + 1} \right) b_3 
\ea
There are four independent parameters to be determined at each temperature $T$ as described in the text.

\end{document}